\documentclass[prb,preprint,showpacs,preprintnumbers,amsmath,amssymb]{revtex4}

\usepackage[dvips]{graphicx}
\usepackage{dcolumn}
\usepackage{bm}
\begin{document}

\preprint{Physical Review B {\bf 73}, 085421 (2006)}

\title{Scanning tunneling microscopy and spectroscopy of the electronic local density of states of graphite surfaces \\near monoatomic step edges}

\author{Y. Niimi}
\author{T. Matsui}
\author{H. Kambara}
\author{K. Tagami}
\author{M. Tsukada}
\author{Hiroshi Fukuyama}
 \email{hiroshi@phys.s.u-tokyo.ac.jp}
\affiliation{%
Department of Physics, University of Tokyo, 7-3-1 Hongo Bunkyo-ku, Tokyo 113-0033, Japan
}%

\date{February 24, 2006}

\begin{abstract}
We measured the electronic local density of states (LDOS) of graphite surfaces near monoatomic step edges, which consist of either the zigzag or armchair edge, with the scanning tunneling microscopy (STM) and spectroscopy (STS) techniques. The STM data reveal that the $(\sqrt{3} \times \sqrt{3}) R 30^{\circ}$ and honeycomb superstructures coexist over a length scale of 3$-$4 nm from both the edges. By comparing with density-functional derived nonorthogonal tight-binding calculations, we show that the coexistence is due to a slight admixing of the two types of edges at the graphite surfaces. In the STS measurements, a clear peak in the LDOS at negative bias voltages from $-100$ to $-20$ mV was observed near the zigzag edges, while such a peak was not observed near the armchair edges. We concluded that this peak corresponds to the graphite ``edge state" theoretically predicted by Fujita \textit{et al.} [J. Phys. Soc. Jpn. {\bf 65}, 1920 (1996)] with a tight-binding model for graphene ribbons. The existence of the edge state only at the zigzag type edge was also confirmed by our first-principles calculations with different edge terminations. 
\end{abstract}

\maketitle

\section{Introduction}
Since the discovery of carbon molecules such as fullerenes~\cite{fullerene} and nanotubes,~\cite{nanotube} \textit{sp$^2$} carbon network systems have been attracting much attention. In these systems, topological structures of the networks critically control the $\pi$ electronic states and material functions. In the case of fullerenes, the relative arrangements of 12 pentagonal rings in the basis of the hexagonal network of carbon atoms are responsible for a variety of electronic properties. Carbon nanotubes further demonstrate that the tubular circumferential (chiral) vector determines whether they are metallic or insulating. In addition to these materials with closed $\pi$ electron networks, nanographites which are nanometer-sized graphite fragments with open edges around the peripheries have novel electronic and magnetic properties,~\cite{enoki1,enoki2} which are not seen in bulk graphite. In the \textit{sp$^2$} network systems with open edges, geometrical arrangements of carbon atoms at the edges should play important roles on the $\pi$ electronic states.

\begin{figure}[htbp]
\begin{center}
\includegraphics[width=10cm]{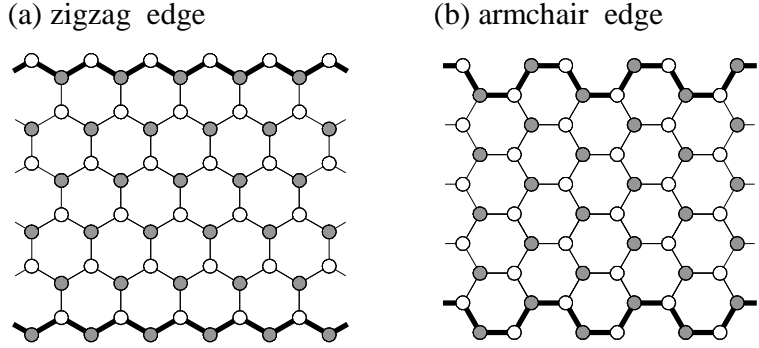}
\caption{Two types of edges for graphite ribbons: (a) zigzag edge and (b) armchair edge. The edges are denoted by the bold lines. The open and closed circles show the B- and A-site carbon atoms, respectively.}
\label{edge_structure_fig}
\end{center}
\end{figure}

Basically, there are two edge shapes in single-layer graphite sheet (graphene), i.e., zigzag and armchair edges (see Fig.~\ref{edge_structure_fig}). Fujita \textit{et al.}~\cite{fujita1,nakada,fujita2} first predicted the existence of the peculiar electronic states localized only at the zigzag edge from the tight binding band calculations for the graphene ribbons. This localized state is known as the graphite``edge state.'' It stems from the topology of the $\pi$ electron networks at the zigzag edge and does not appear at the armchair edge. The flat band nature of the edge state results in a peak in the local density of states (LDOS) at the Fermi energy ($E_{F}$). When the ribbon width is large enough, the influence of the edge state on the total density of states is negligible. However, the LDOS near the zigzag edge is strongly affected by the edge state, which would be observable with the scanning tunneling spectroscopy (STS) technique. A similar edge state is also obtained for multilayer ribbons of $\alpha \beta$ stacking from the first-principles calculations.~\cite{miyamoto} This indicates that the edge state would exist in more realistic systems, such as step edges at bulk graphite surfaces.

In the previous STS measurements performed near circular edges of graphite nanopits,~\cite{klusek1,klusek2} a broad maximum was observed in the LDOS near $E_{F}$, which was attributed to the edge state. However, it is rather difficult, in this case, to distinguish between the electronic properties of the zigzag and armchair edges because both types of the edges inevitably coexist nearby on almost equal footing in the circular edge. Scanning tunneling microscopy (STM) measurements have been performed at linear step edges at the surface of highly oriented pyrolytic graphite (HOPG).~\cite{armchair_stm,zigzag_stm} So far, the $(\sqrt{3} \times \sqrt{3}) R 30^{\circ}$ superstructure has been clearly observed near the armchair edges,~\cite{armchair_stm} while it is not so clear yet whether similar superstructures exist near the zigzag edges.~\cite{armchair_stm,zigzag_stm}

Here, we investigated the LDOS in the vinicity of linear step edges of both the zigzag and armchair types with STM/STS. A Brief Report of the STS observation of the graphite edge state has been already made elsewhere.~\cite{ass} In this paper, we present more detailed STM/STS data and more advanced theoretical analyses of the edge state. In the following section, experimental details including sample preparations are described. In Sec. III A, we show typical STM images of graphite surfaces in large area and higher resolution images in the vinicity of both the zigzag and armchair edges. The STS data near both the edges are presented in Sec. III B. Section IV is devoted to discuss the experimental results and theoretical models to be compared with them.

\section{Experimental details}
We used two kinds of graphites, i.e., HOPG and \textit{ZYX} exfoliated graphite~\cite{niimi} (hereafter \textit{ZYX}) to find linear step edges. HOPG is synthesized by chemical vapor deposition and subsequent heat treatment under high pressures. It is polycrystalline graphite with ordered $c$-axis orientation (the rocking angle $\theta \leq 0.5^{\circ}$). The \textit{ZYX} samples were made from HOPG by the graphite intercalation technique with HNO$_3$ and by subsequent evacuation of the intercalant at 600 $^{\circ}$C. And then, it was heated at 1500 $^{\circ}$C for 3 h to remove the remnant intercalants. \textit{ZYX} is primarily used as an adsorption substrate for studies of monolayer atoms~\cite{helium,xenon,birgeneau} and molecules,~\cite{hydrogen} due to its large specific surface area and moderately large single crystalline (platelet) size. It is reported in the previous scattering experiments~\cite{birgeneau} that \textit{ZYX} has a platelet size of 100$-$200 nm, which is an order of magnitude smaller than HOPG. Thus, the step edges should be more easily found in \textit{ZYX} compared to HOPG. Other characteristics of \textit{ZYX} have been published elsewhere.~\cite{niimi} All graphite edges studied here are monoatomic in height with almost linear shape over the length scale of several tens nanometers. We believe that active $\sigma$-orbital bonds at the edges are terminated by hydrogen or else in air since we did not remove them intentionaly in ultrahigh vacuum (UHV) at elevated temperatures.

The STM/STS measurements were carried out with homemade STMs.~\cite{ULT-STM} The STM images were taken at room temperature in air with a tunnel current ($I$) of 1.0 nA and a typical bias voltage ($V$) of $+0.05$ to $+1.0$ V in the constant current mode. Mechanically sharpened Pt$_{0.8}$Ir$_{0.2}$ and electrochemically etched W wires were used as STM tips. The STS measuremets were performed at $T = 77$ K in UHV ($P \leq 2 \times 10^{-7}$ Pa). A tunnel spectrum was obtained by averaging a set of $dI/dV$ vs. $V$ curves measured with the lock-in technique ($f = 71.73$ or 412 Hz, $V_{{\rm mod}} = 1$ or 6 mV) at 100 to 900 grid points over $5 \times 5$ to $15 \times 15$ nm$^2$ area. All the results shown here did not depend on the tip material and lock-in parameter.

\section{Results}
\subsection{STM observations of graphite edges}
\begin{figure}[htbp]
\begin{center}
\includegraphics[width=12cm]{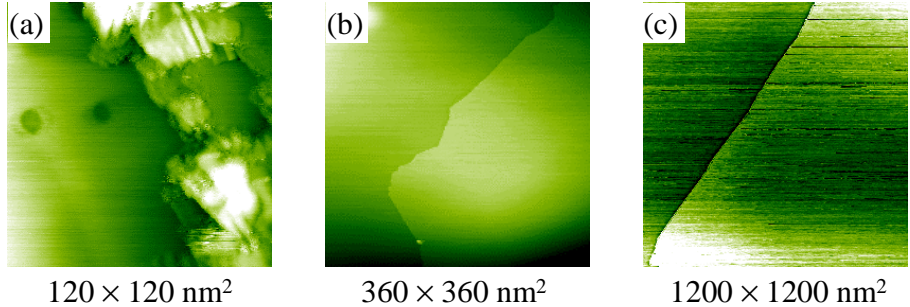}
\caption{Typical STM images of three kinds of graphites ($T=300$ K, in air, $I=1.0$ nA, $V=0.5$ V): (a) Grafoil ($120 \times 120$ nm$^{2}$), (b) \textit{ZYX} ($360 \times 360$ nm$^{2}$), and (c) HOPG ($1200 \times 1200$ nm$^{2}$).}
\label{stm_graphite_fig}
\end{center}
\end{figure}

Figure~\ref{stm_graphite_fig} shows exemplary STM images of three kinds of graphites [(a) Grafoil,~\cite{grafoil} (b) \textit{ZYX}, (c) HOPG]. They show that moderately long linear steps ($\geq 100$ nm) are available at the surfaces of \textit{ZYX} and HOPG. The featureless areas are atomically flat. From STM images taken over a wider area range, we determined the platelet size distributions of \textit{ZYX} and Grafoil. The results are shown in the Appendix.

\begin{figure}[htbp]
\begin{center}
\includegraphics[width=10cm]{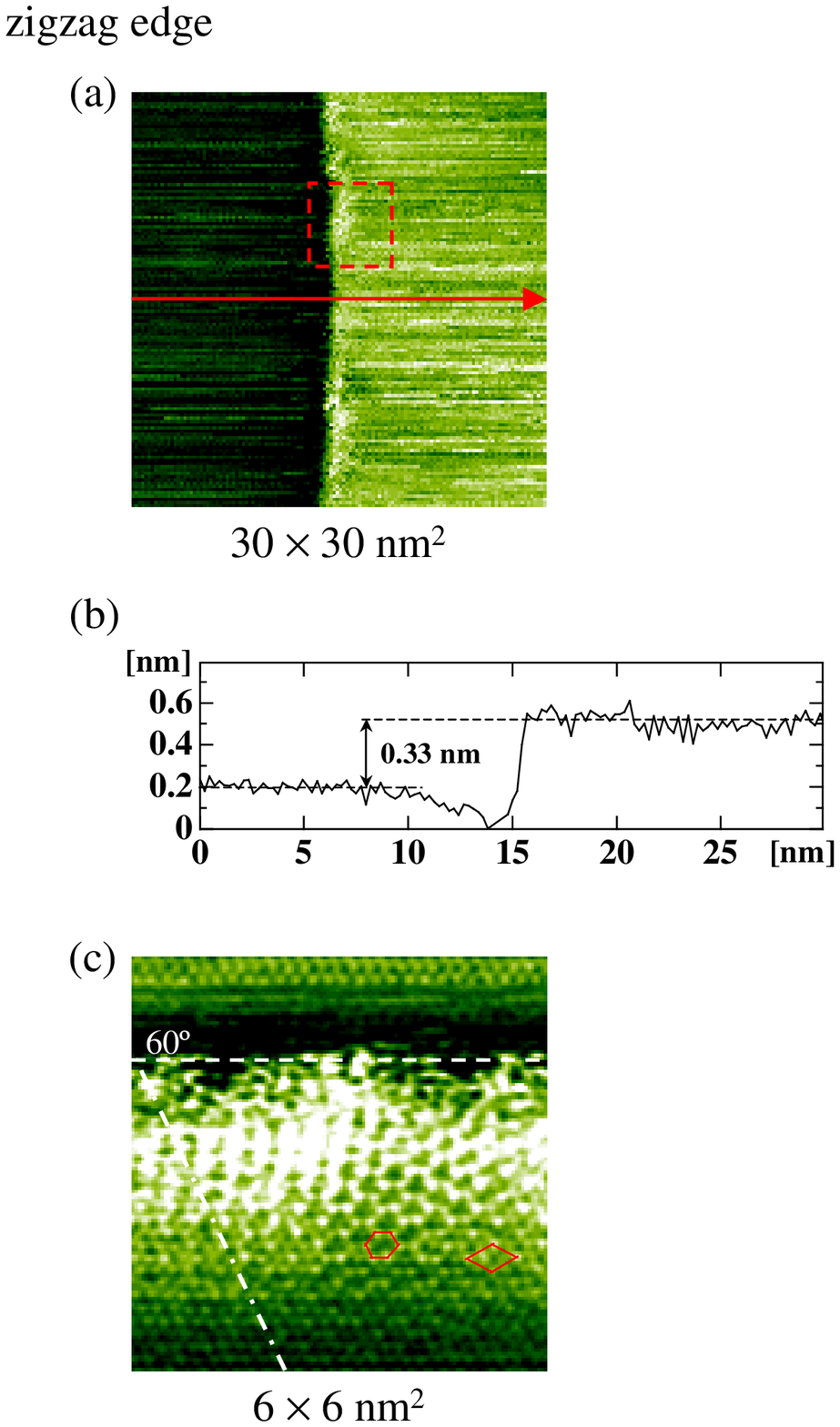}
\caption{STM images and a cross section near a monoatomic step with zigzag edge at the surface of \textit{ZYX} ($T=300$ K, in air, $I=1.0$ nA, $V=0.1$ V). (a) $30 \times 30$ nm$^{2}$ scan image. (b) Cross section profile along the arrow in (a). (c) $6 \times 6$ nm$^{2}$ scan image of the square region in (a). The dashed and dot-dashed lines show the edge and the atomic row of B-site atoms, respectively. The diamond and hexagon represent the $(\sqrt{3} \times \sqrt{3}) R 30^{\circ}$ superstructure and honeycomb one.}
\label{zigzag_superstructure_fig}
\end{center}
\end{figure}

In Fig.~\ref{zigzag_superstructure_fig}(a), we present an STM image obtained near a zigzag edge at the surface of \textit{ZYX}. The step edge looks extending straight over 30 nm. The step height estimated from line profile analysis is 0.33 nm [Fig.~\ref{zigzag_superstructure_fig}(b)], which corresponds to the layer spacing of graphite (0.335 nm). In Fig.~\ref{zigzag_superstructure_fig}(c), we show a higher resolution image of the square region denoted in Fig.~\ref{zigzag_superstructure_fig}(a). The scan direction is rotated here by $90^{\circ}$ with respect to that in Fig.~\ref{zigzag_superstructure_fig}(a). Although a good atomic resolution is not obtained right on the edge, it can be identified as zigzag type since the atomic row of B-site carbon atoms (the dot-dashed line) is oriented at 60$^{\circ}$ to the edge direction [see Fig.~\ref{edge_structure_fig}(a)]. Note that the edge shown here is probably not a pure zigzag edge but that mingled with a small fraction of armchair edges. Large electronic density of states is observed within 2 nm from the edge. Moreover, two types of superstructures coexist only on the upper terrace. One is the $(\sqrt{3} \times \sqrt{3}) R 30^{\circ}$ superstructure, and the other is the honeycomb one which consists of six B-site carbon atoms. These superstructures extend over 3$-$4 nm from the edge. The superstructure pattern does not depend on the bias voltage in a range between $+0.05$ and $+1.0$ V. Such superstructures were also obtained at the HOPG surface at 77 K in UHV.

\begin{figure}[htbp]
\begin{center}
\includegraphics[width=10cm]{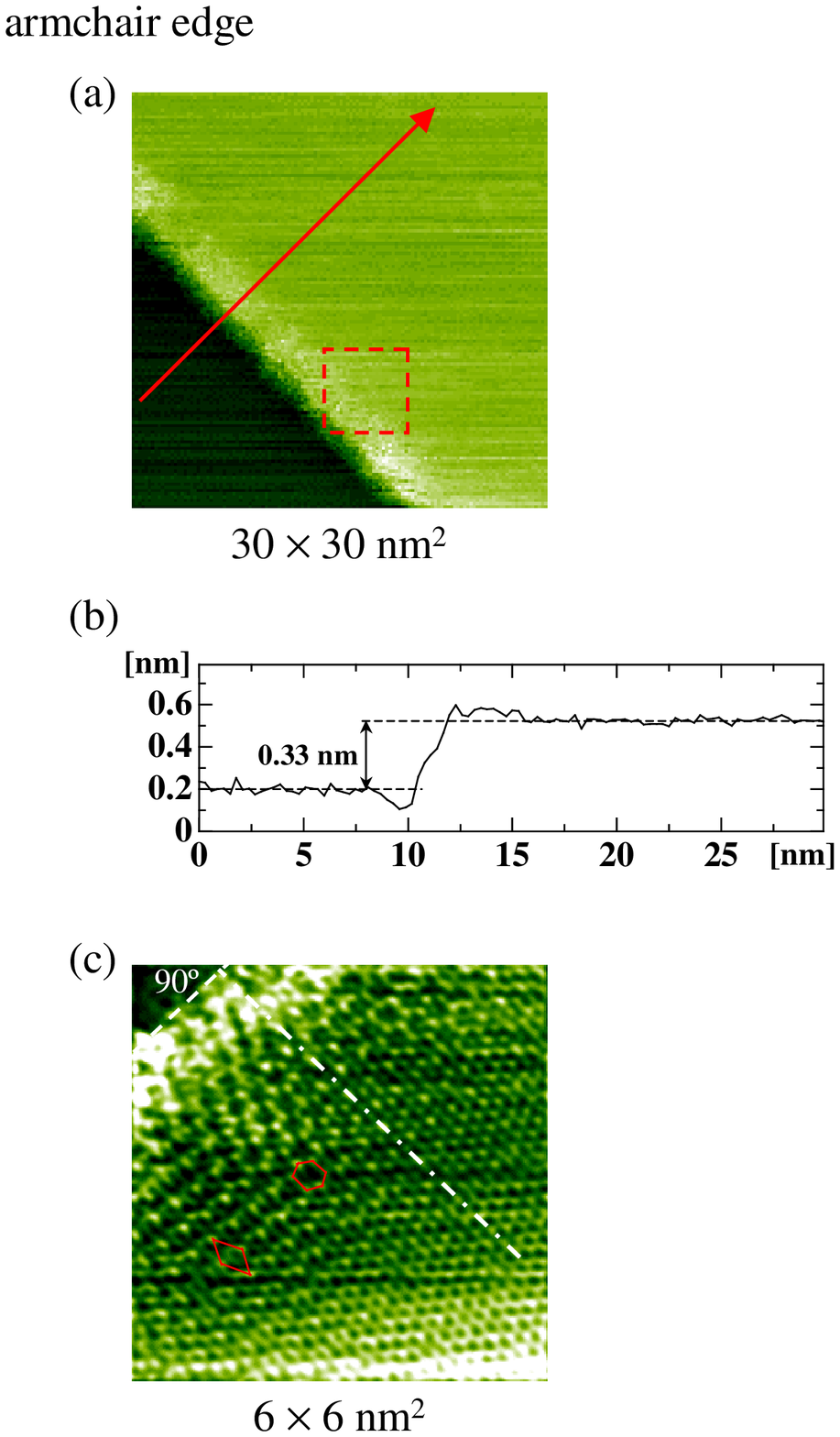}
\caption{STM images and a cross section near a monoatomic step with armchair edge at the \textit{ZYX} surface ($T=300$ K, in air, $I=1.0$ nA, $V=0.1$ V). (a) $30 \times 30$ nm$^{2}$ scan image. (b) Cross section profile along the arrow in (a). (c) $6 \times 6$ nm$^{2}$ scan image of the square region in (a). The dashed line, dot-dashed line, diamond, and hexagon represent the same meanings as in Fig.~\ref{zigzag_superstructure_fig}(c).}
\label{armchair_superstructure_fig}
\end{center}
\end{figure}

In Fig.~\ref{armchair_superstructure_fig}, we show STM images near an armchair edge at the \textit{ZYX} surface. The edge is a monoatomic step edge extending straight over 30 nm [Figs.~\ref{armchair_superstructure_fig}(a) and 4(b)]. It is identified as armchair type from the fact that the atomic row of B-site carbon atoms [the dot-dashed line in Fig.~\ref{armchair_superstructure_fig}(c)] is oriented at 90$^{\circ}$ to the edge direction [see Fig.~\ref{edge_structure_fig}(b)]. As in the case of zigzag edge, both the $(\sqrt{3} \times \sqrt{3}) R 30^{\circ}$ and honeycomb superstructures are observed extending over 3$-$4 nm from the edge. The similar coexistence of the superstructures has been reported by Giunta and Kelty~\cite{armchair_stm} for an armchair step edge at HOPG surface.

\subsection{STS observations of graphite edges}

\begin{figure*}[htbp]
\begin{center}
\includegraphics[width=12cm]{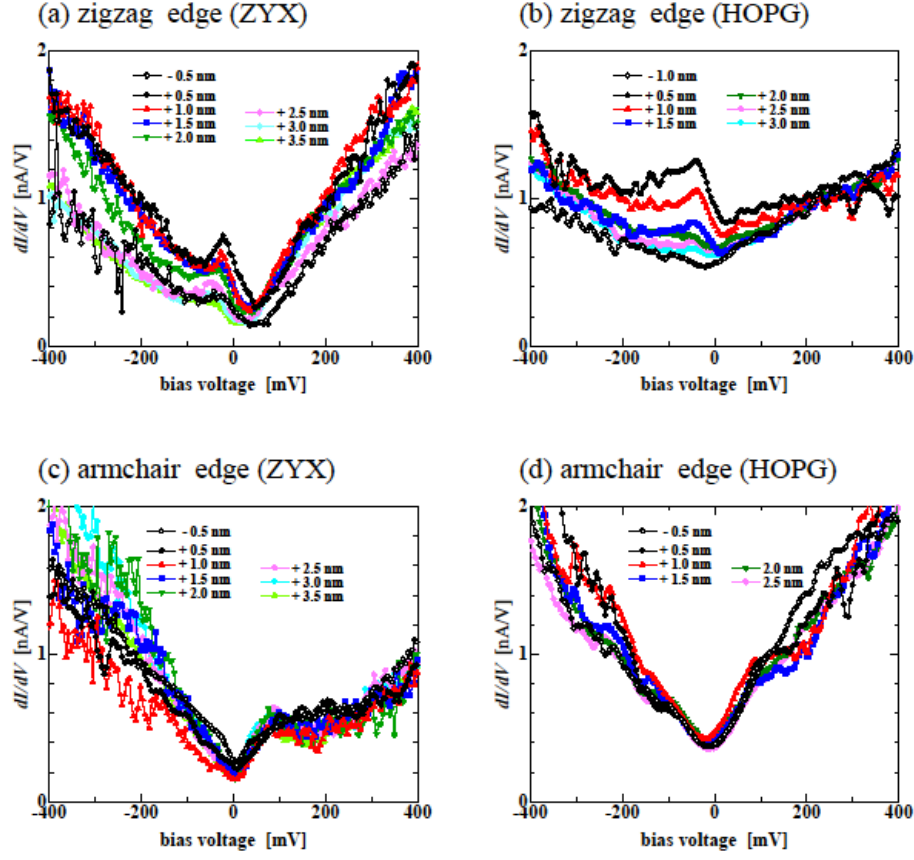}
\caption{$dI/dV$ curves measured at $|V|\leq 0.4$ V near zigzag edges and armchair edges at the surfaces of \textit{ZYX} [(a),(c)] and HOPG [(b),(d)] ($T=77$ K, in UHV). The numbers denoted are distances from the edge ($d=0$).}
\label{zigzag_armchair_sts1_fig}
\end{center}
\end{figure*}

STS data in the vinicity of single step edges at the \textit{ZYX} and HOPG surfaces were obtained at 77 K in UHV. The scan directions were fixed parallel to the edges. Figures~\ref{zigzag_armchair_sts1_fig}(a) and 5(b) show tunnel spectra measured in the bias voltage range of $|V|\leq 0.4$ V near zigzag edges at the surfaces of \textit{ZYX} and HOPG, respectively. In order to obtain better signal-to-noise ratios, 8 to 30 $dI/dV$ curves taken at fixed distances ($d$) from the edge were averaged. A clear peak appears at negative bias voltages from $-100$ to $-20$ mV for $0 < d < 3$ nm. It grows as the tip approaches the edges on the terrace ($d > 0$) but suddenly disappears when it moves across the edges ($d < 0$). Such behavior does not depend on graphite sample (\textit{ZYX} or HOPG). Since the tunnel current was unstable at $|d| < 0.5$ nm for some reason, we could not obtain reliable spectra right on the edge. It should also be noted that the definition of $d = 0$ is somewhat arbitrary ($\pm 0.5$ nm) because of the degraded spatial resolution in that region.

By contrast with the case of zigzag edge, we obtained qualitatively different spectra near armchair edges. As is shown in Figs.~\ref{zigzag_armchair_sts1_fig}(c) and 5(d), the tunnel spectra near the armchair edges do not have such a peak within the experimental errors. These spectra are essentially independent of $d$. Therefore, the LDOS peak depending on $d$ in Figs.~\ref{zigzag_armchair_sts1_fig}(a) and 5(b) should correspond to the graphite edge state that has been theoretically predicted to exist only for the zigzag edge.~\cite{fujita1,nakada,fujita2,miyamoto} In Fig.~\ref{zigzag_armchair_sts1_fig}(c), there is a bump like structure at positive voltages ($0.08 \leq V \leq 0.1$ V). It is probably due to a local electrostatic potential induced by the tip. In STS experiments at graphite surfaces,~\cite{matsui} we often observed similar LDOS bumps depending on the tip conditions but not on the spatial position.

It is claimed in the previous STS measurements near the circular edges~\cite{klusek1,klusek2} that the LDOS peak of 0.2 eV wide which appears in the positive energy range of 0.02$-$0.25 eV as $d \to 0$ should correspond to the edge state. However, these results are not consistent with our results in terms of peak energy and width. Although we do not know the reason for the discrepancy between the previous data and ours, the complicated structure of the circular edges might be responsible for that.

\begin{figure}[htbp]
\begin{center}
\includegraphics[width=7cm]{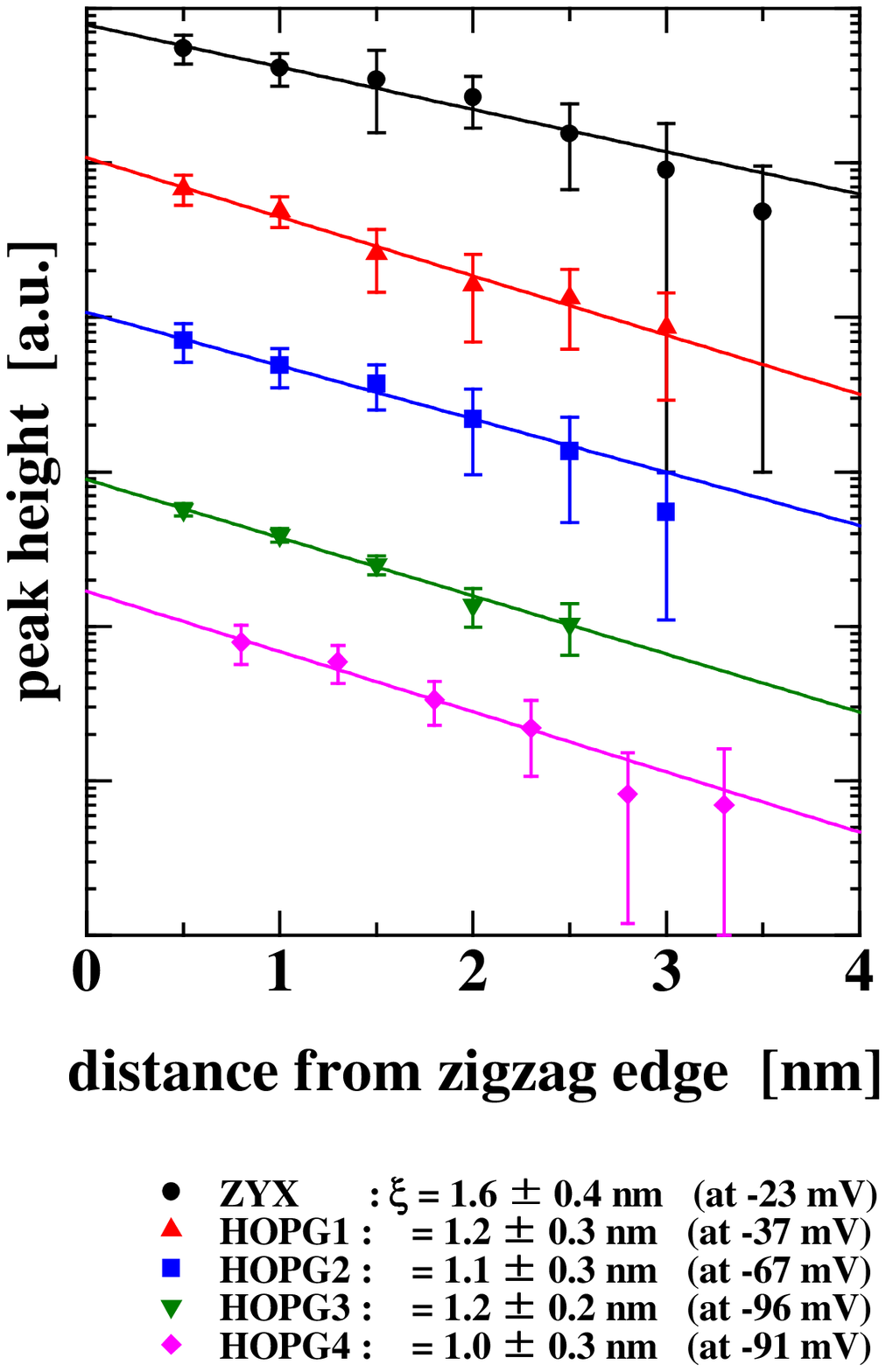}
\caption{Semilog plot of the distance ($d$) dependece of the peak heights at five zigzag edges associated with the graphite edge state. Each plot is vertically shifted for clarity. The lines show exponential fittings of the data.}
\label{peak_height_fig}
\end{center}
\end{figure}

Figure~\ref{peak_height_fig} is a semilog plot of the $dI/dV$ peak heights observed near five different zigzag edges as a function of $d$. Note that the peak heights shown here were obtained by subtracting a smooth background from the raw spectra. From this plot, we can determine a decay length ($\xi$) of the localized state. The averaged $\xi$ was estimated as $1.2 \pm 0.3$ nm.

\begin{figure}[htbp]
\begin{center}
\includegraphics[width=6cm]{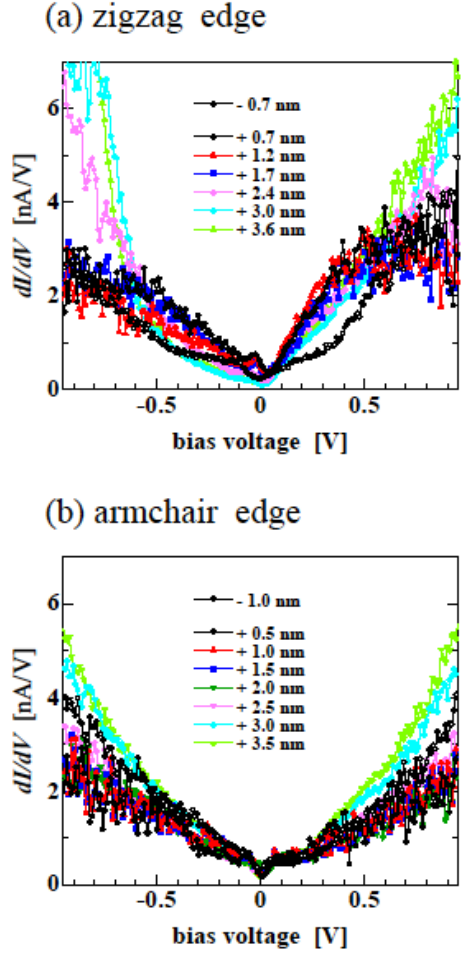}
\caption{$dI/dV$ curves measured at $|V|\leq 1.0$ V near (a) a zigzag edge and (b) an armchair edge at the \textit{ZYX} surface ($T=77$ K, in UHV). The numbers denoted are distances from the edge ($d=0$).}
\label{zigzag_armchair_sts2_fig}
\end{center}
\end{figure}

In a wide bias voltage range between $-1.0$ and $+1.0$ V, other interesting properties are found. The spectra obtained far away ($d \geq 3.5$ nm) from the edges are ``V-shaped'' ones, which are characteristic of graphite. However, the LDOS decreases selectively in a voltage range of $|V|\geq 0.6$ V within a distance of 2 nm from the zigzag edge [Fig.~\ref{zigzag_armchair_sts2_fig}(a)]. A similar decrease has been observed near the circular edges.~\cite{klusek1} On the other hand, the LDOS decreases in the whole voltage range for $d < 2.5$ nm near the armchair edge [Fig.~\ref{zigzag_armchair_sts2_fig}(b)].

\section{Theoretical simulations and discussions}
\subsection{Two types of superstructures}
In this section, we discuss the origin of the two types of superstructures observed in the experimental STM images near both the zigzag and armchair edges. These superstructures have been also observed by other authors in the vinicity of defects on HOPG surfaces such as grain boundaries,~\cite{grain_boundaries} deposited metal particles,~\cite{adsorbed_metal1,adsorbed_metal2,adsorbed_metal3,adsorbed_metal4,adsorbed_metal5} or holes made by spattering.~\cite{hole1,hole2} They are attributed to an interference between incident and scattered electron wave functions.~\cite{adsorbed_metal1} However, topology of the defect sites was not known in these experiments, while the experimental results were explained by arbitrary combinations between the incident and scattered wave functions.~\cite{adsorbed_metal3,adsorbed_metal4} On the other hand, the zigzag and armchair edges have well-defined structures in atomic scale. In the systems with these edges, the pattern and periodicity of the superstructures should be obtained without any assumptions. We theoretically analyzed the local electronic states near both the zigzag and armchair edges and compared the calculated results with the STM images described in the previous section. 

Our simulations were made on a double-layer graphene system with the zigzag or armchair edge, which is more realistic than the graphene ribbon or multilayer ribbons. The bottom layer is an infinite graphene without edges. The top layer consists of periodically arranged graphene ribbons with the zigzag or armchair edge (zigzag or armchair ribbons), whose widths are 15.7 and 8.6 nm, respectively. The spacing between the ribbons is long enough, and the periodic boundary conditions are imposed along the ribbon directions. The electronic states of these graphite layers were calculated by the density-functional derived nonorthogonal tight-binding model.~\cite{Frauenheim98} We assumed that carbon atoms at the edges are hydrogen terminated, and took into account only the $\pi$ orbital at each carbon site which is relevant to the electronic states near $E_{F}$. The LDOS at each atomic site was obtained by diagonalizing the Hamiltonian and overlap matrices assosicated with the $\pi$ orbitals at the $\Gamma$ point.

\begin{figure}[htbp]
\begin{center}
\includegraphics[width=6cm]{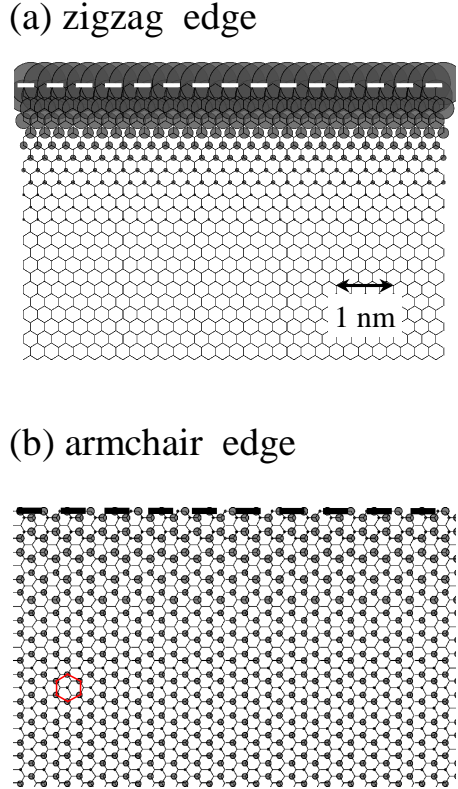}
\caption{Simulations for tunnel currents at the B-sites near (a) the perfect zigzag and (b) armchair edges by the nonorthogonal tight-binding model. The radii of the circles on the sites indicate integrals of the calculated LDOS in an energy range between $E_{F}$ and $+0.1$ eV. The white and black dashed lines represent the zigzag and armchair edge, respectively.}
\label{perfect_edge_cal_fig}
\end{center}
\end{figure}

Figure~\ref{perfect_edge_cal_fig} shows spatial variations of calculated electronic states in the vinicity of the zigzag and armchair edges. The radii of the circles plotted on the B-sites in the figure represent integrals ($I_{\rm cal}$) of the calculated LDOS over an energy range between $E_{F}$ and $+0.1$ eV. Note that $I_{\rm cal}$ corresponds to the local tunnel currents at $V=+0.1$ V in the experimental STM images. Figure~\ref{perfect_edge_cal_fig}(a) indicates the existence of the localized electronic state in the vinicity of the zigzag edge with $\xi \sim 0.5$ nm. In this system, there appear no superstructures anywhere. Conversely, Fig.~\ref{perfect_edge_cal_fig}(b) does not show such localized states near the armchair edge but a honeycomb superstructure persisting far beyond 5 nm from the edge. These calculated results for the perfect edges are inconsistent with our experimental observations.

\begin{figure}[htbp]
\begin{center}
\includegraphics[width=6cm]{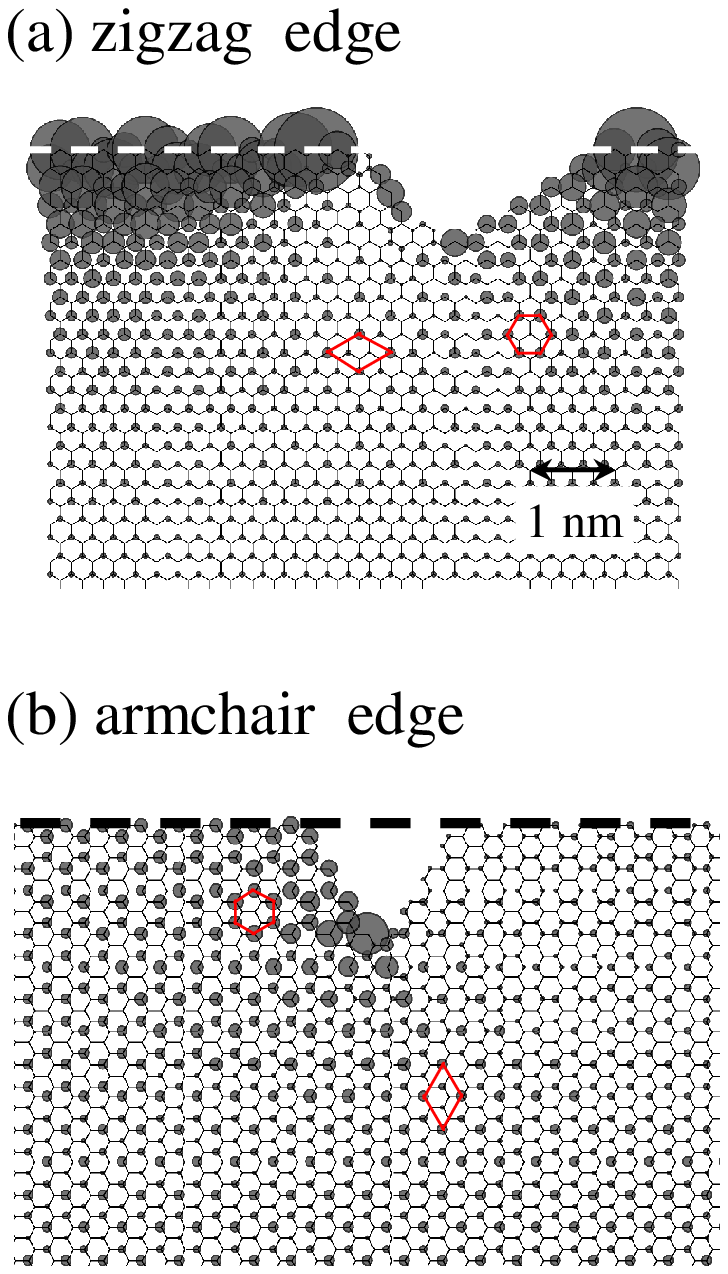}
\caption{Simulations for tunnel currents at the B-sites near (a) a zigzag edge with a small amount of armchair edges and (b) an armchair edge with a small amount of zigzag edges by the non-orthogonal tight-binding model.}
\label{mixture_edge_cal_fig}
\end{center}
\end{figure}

Hence, we have calculated the LDOS near the zigzag (armchair) edges which are mingled with small amounts of armchair (zigzag) edges. We show two examples for such edge patterns in Figs.~\ref{mixture_edge_cal_fig}(a) and 9(b). In this case, both the $(\sqrt{3} \times \sqrt{3}) R 30^{\circ}$ and honeycomb superstructures appear on the terrace with the zigzag edges slightly admixed with armchair edges. The superstructures extend over 4$-$5 nm from the edge and have complicated distributions in the parallel direction to the edge. In spite of admixing of armchair edges, the localized state still remains near the zigzag edge, but its decay length becomes longer ($\xi \sim 1.2$ nm) than that for the perfect zigzag edge ($\xi \sim 0.5$ nm). This calculation reproduces fairly well the spacial extensions of the two types of superstructures and the localized state observed in the experiment [see Figs.~\ref{zigzag_superstructure_fig}(c) and~\ref{peak_height_fig}]. Unfortunately, we could not observe the atomic arrangement of the zigzag edge clearly in the experiment. Such observations will be done in future works. Nevertheless, the spacial distributions of the two types of superstructures and the localized state strongly indicate that the edge in Fig.~\ref{zigzag_superstructure_fig}(c) is mingled with a small amount of armchair edges.

The coexistence of the two types of superstructures is also seen in the calculation for the armchair edges slightly admixed with zigzag edges in Fig.~\ref{mixture_edge_cal_fig}(b). This again reproduces the feature of the experimental image in Fig.~\ref{armchair_superstructure_fig}(c). However, the experimental spatial extensions of the superstructures (3$-$4 nm) are shorter than the calculated ones (far beyond 5 nm). This may be due to the three dimensional character of the experimental system, which is not fully taken into account in the present calculations. 

\subsection{Graphite edge state}

In order to examine theoretically the detailed features of LDOS near the step edges around $E_F$ ($ | E | \leq 0.2$ eV), we performed the first-principles calculations based on the density functional theory with the generalized gradient approximation. We adopt the exchange-correlation potential introduced by Perdew \textit{et al}.~\cite{Perdew96} The cutoff of the plane-wave basis set is assumed to be 20.25 Ry. The calculations were performed on the double-layer graphene system as denoted in the previous section. The zigzag and armchair ribbon widths of the top layer are 1.565 and 0.862 nm, respectively. The height of the ribbons is 0.1189 nm for both cases. The distance between the top layer and the infinite bottom layer is 0.3356 nm. The edge carbon atoms are terminated by H atoms or OH groups. The lengths between C-C, C-H, C-O, O-H bonds are fixed to be 0.14226, 0.110, 0.140, 0.100 nm, respectively. We adopt the Vanderbilt type ultrasoft pseudopotentials~\cite{Vander} for C, H, and O atoms. The theoretical LDOS shown below is obtained by integrating the LDOS calculated at the lattice point over the volume of each atom.

\begin{figure*}[htbp]
\begin{center}
\includegraphics[width=15cm]{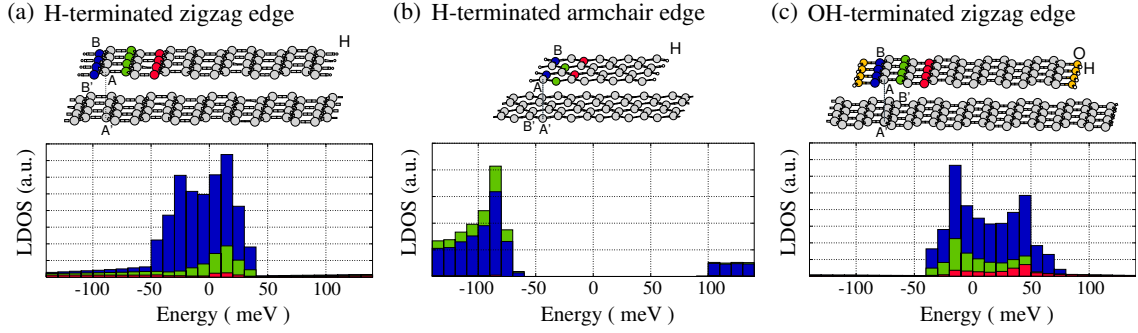}
\caption{First-principles calculations of LDOS for different edges and terminations. The different colors correspond to different sites denoted in the sketches of the double-layer graphene system on which the calculations were performed. (a) H-terminated zigzag edge, (b) H-terminated armchair edge, and (c) OH-terminated zigzag edge.}
\label{theoretical_ldos_fig}
\end{center}
\end{figure*}

Figures~\ref{theoretical_ldos_fig}(a) and 10(b) show the calculated LDOS at the B-sites for the zigzag and armchair ribbons whose edges are H terminated. In the former case, the LDOS peak due to the edge state appears near $E_F$ at $d=0$, and rapidly decreases with increasing $d$. The peak width of the theoretical LDOS at $d=0$ is about 80 meV, which is consistent with the experimental ones at $d=0.5$ nm [see Figs.~\ref{zigzag_armchair_sts1_fig}(a) and 5(b)]. Note again that the definition of $d = 0$ in this experiment is somewhat arbitrary ($\pm 0.5$ nm). On the other hand, such a peak dose not appear in the armchair case. Therefore, we conclude that the LDOS peak experimentally observed just below $E_F$ in Figs.~\ref{zigzag_armchair_sts1_fig}(a) and 5(b) originates from the graphite edge state. Although the experimental edge state appears at negative energies ($-100$ to $-20$ mV), the theoretical one obtained in the first-principles calculations does at a slightly higher energy which is more or less close to $E_{F}$. Recently, Sasaki \textit{et al.}~\cite{energy_shift_theory} have shown that within the tight binding approximation, the edge state shifts to the negative side ($E < 0$) if the next-nearest-neighbor hopping process is taken into account.

Next, we discuss the influence of terminated atoms and/or molecules on the edge state. The graphite edge state is originally predicted for the H-terminated graphene ribbons.~\cite{fujita1,nakada,fujita2} Since the edges observed in our STS measurements had been exposed in air before being loaded into the UHV chamber, it is possbile that the edges are terminated either by hydrogen or hydroxide. We thus performed the same first-principles calculations for an OH-terminated zigzag edge as well. In Fig.~\ref{theoretical_ldos_fig}(c), we show the calculated LDOS for the OH-terminated zigzag ribbons on the infinite graphene. Although the peak width is slightly larger than that for the H-terminated zigzag ribbons, the LDOS peak still remains near $E_{F}$. Therefore, the LDOS peak assosiated with the graphite edge state is not strongly affected by the different terminations in air, i.e., H atom or OH group. Recently, Kobayashi \textit{et al.}~\cite{h-terminated} have made STM/STS measurements near the zigzag and armchair edges with H and ambient terminations. They found a similar LDOS peak for the zigzag edge with H temination to that obtained in this work in terms of width and energy. Note that the $d$ dependence of the peak was not studied systematically in their experiment.

We observed the two characteristically different decreases of the LDOS at larger magnitudes of voltages on approaching the zigzag and armchair edges, as discussed in the last paragraph of Sec. III B. Near the zigzag edge, we observed that the LDOS is suppressed selectively at $|V| \geq$ 0.6 V. Kulsek \textit{et al.}~\cite{klusek1,klusek2,klusek3} observed the similar LDOS suppression near the circular edges at $|V| \geq$ 0.6$-$0.8 V, and claimed that it is associated with the $\pi$ band splitting due to multilayer interaction at the $P$ point in the two-dimensional Brillouin zone.~\cite{splitting1,splitting2} However, it is not clear why such suppression becomes prominent with dicreasing $d$ with this explanation. We have reasonable explanations neither for the other type of LDOS decrease (at $|V| \geq$ 0.3 V) near the armchair edge at this moment.

\section{Conclusions}
We have studied the electronic local density of states (LDOS) near single step edges at graphite surfaces. In scanning tunneling microscopy measurements, the $(\sqrt{3} \times \sqrt{3}) R 30^{\circ}$ and honeycomb superstructures were observed over 3$-$4 nm from both the zigzag and armchair edges. Calculations based on a density-functional derived nonorthogonal tight binding model show that admixing of the two types of edges is responsible for the experimental coexistence of these superstructures. Scanning tunneling spectroscopy measurements near the zigzag edges reveal the existence of a clear peak in the LDOS at several tens meV below the Fermi energy. The peak amplitude grows as we approach the edge on the terrace, but suddenly diminishes across the edge. No such a peak was observed near the armchair edges. The first-principles calculations for the zigzag and armchair ribbons on infinite graphene sheets reproduce well these experimental results. Therefore, we conclude that the LDOS peak experimentally observed only at the zigzag edge corresponds to the graphite edge state theoretically predicted in the previous calculations on the graphene ribbons. The decay length of the edge state in this experiment is about 1.2 nm, which is consistent with calculations for a zigzag edge slightly mingled with armchair edges.

\begin{acknowledgments}
One of us (H.F.) thanks the late Mitsutaka Fujita for stimulating his interest in the graphite edge state. The authors are grateful to H. Akisato for useful comments on this manuscript. This work was financially supported by Grant-in-Aid for Scientific Research from MEXT, Japan and ERATO Project of JST. Y.N. and T.M. acknowledge the JSPS for financial support.
\end{acknowledgments}

\appendix*

\section{Platelet size distributions of exfoliated graphites}
\begin{figure}[htbp]
\begin{center}
\includegraphics[width=7cm]{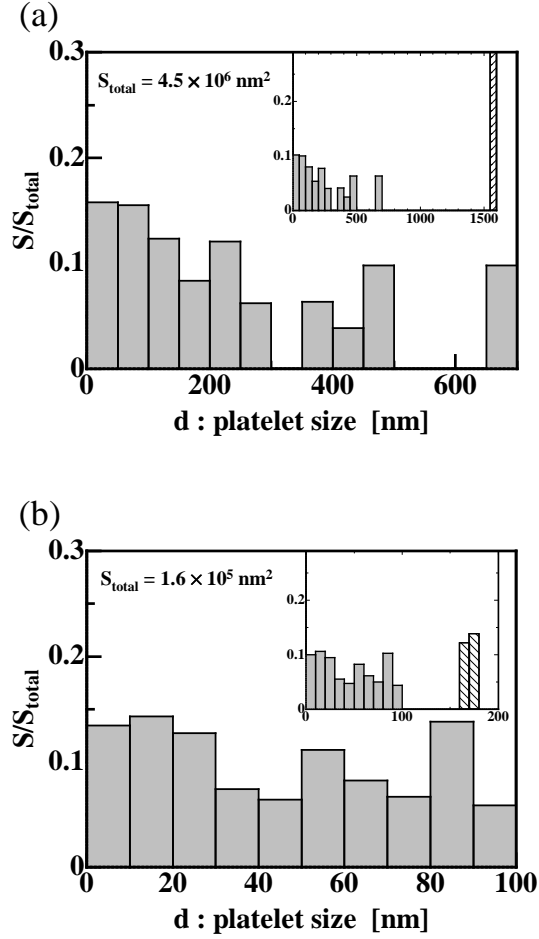}
\caption{Platelet size distributions of (a) \textit{ZYX} and (b) Grafoil. The insets show the distributions in wider sizes. The hatched data correspond to regions where normal intercalation and exfoliation processes have not been taken place.
}
\label{histogram_all_fig}
\end{center}
\end{figure}

In this Appendix, we present results of STM measurements on the platelet size distributions of exfoliated graphites. Figures~\ref{histogram_all_fig}(a) and 11(b) are histograms showing such distributions of \textit{ZYX} and Grafoil, respectively. Here we define the platelet size as the square root of measured platelet area. In the inset of Fig.~\ref{histogram_all_fig}(a), the distribution above 1500 nm which is well separated from that below 700 nm is probably a contribution from surface regions where normal intercalation and exfoliation processes have not taken place.~\cite{niimi} Such a contribution is also seen in the inset of Fig.~\ref{histogram_all_fig}(b). Apart from these contributions, both the platelet size distributions are relatively wide and featureless. The average platelet sizes of \textit{ZYX} and Grafoil, which are weighted with areas, are estimated as 240 and 45 nm, respectively. These values are consistent with those estimated in the previous scattering experiments.~\cite{birgeneau}

\newpage 

\end{document}